%% file: main.tex
\documentclass[english,submission,copyright,creativecommons]{eptcs}
\usepackage[T1]{fontenc}
\usepackage[latin9]{inputenc}
\usepackage{array}
\usepackage{multirow}
\usepackage{graphicx}
\usepackage{eurosym}
\usepackage{threeparttable}



\usepackage{babel}

\author{Muhammad Sabir Idrees
\institute{Institut Mines-T\'{e}l\'{e}com, T\'{e}l\'{e}com Bretagne,\\
  2, rue de la Ch\^ataigneraie, 35512 Cesson S\'evign\'e, France}
\email{muhammad.idrees@telecom-bretagne.eu}
\and
Yves Roudier
\institute{Institut EURECOM,\\
2229 Route des Cr\^etes,\\ Sophia Antipolis, 06560 Valbonne, France}
\email{Yves.Roudier@eurecom.fr}
\and
Ludovic Apvrille
\institute{Institut Mines-T\'{e}l\'{e}com,  T\'{e}l\'{e}com ParisTech,\\
  46 rue Barrault, 75634 Paris, France}
\email{ludovic.apvrille@telecom-paristech.fr}
}

\def\titlerunning{Model the System from Adversary Viewpoint}
\def\authorrunning{M. Sabir Idrees, Y. Roudier, L. Apvrille}

\begin{document}

\title{Model the System from Adversary Viewpoint: Threats Identification and Modeling}
\maketitle
\begin{abstract}
Security attacks are hard to understand, often expressed with
unfriendly and limited details, making it difficult for security
experts and for security analysts to create intelligible security
specifications. For instance, to explain {\it Why} (attack objective),
{\it What} (i.e., system assets, goals, etc.), and {\it How} (attack
method), adversary achieved his attack goals. We introduce in this
paper a security attack meta-model for our SysML-Sec framework
\cite{EURECOM+4119}, developed to improve the threat identification
and modeling through the explicit representation of security concerns
with knowledge representation techniques. Our proposed meta-model
enables the specification of these concerns through ontological
concepts which define the semantics of the security artifacts and
introduced using SysML-Sec diagrams. This meta-model also enables
representing the relationships that tie several such concepts
together. This representation is then used for reasoning about the
knowledge introduced by system designers as well as security experts
through the graphical environment of the SysML-Sec framework.
\end{abstract}

\input{Introduction}
\label{sec:Introduction}
\section{Collecting Knowledge about Adversary}
\label{sec:MSP}
\input{MSP}

\section{Security Ontology}
\label{subsec:so}

\begin{figure}[http] \centering
   \includegraphics [width=4in, height=3in] {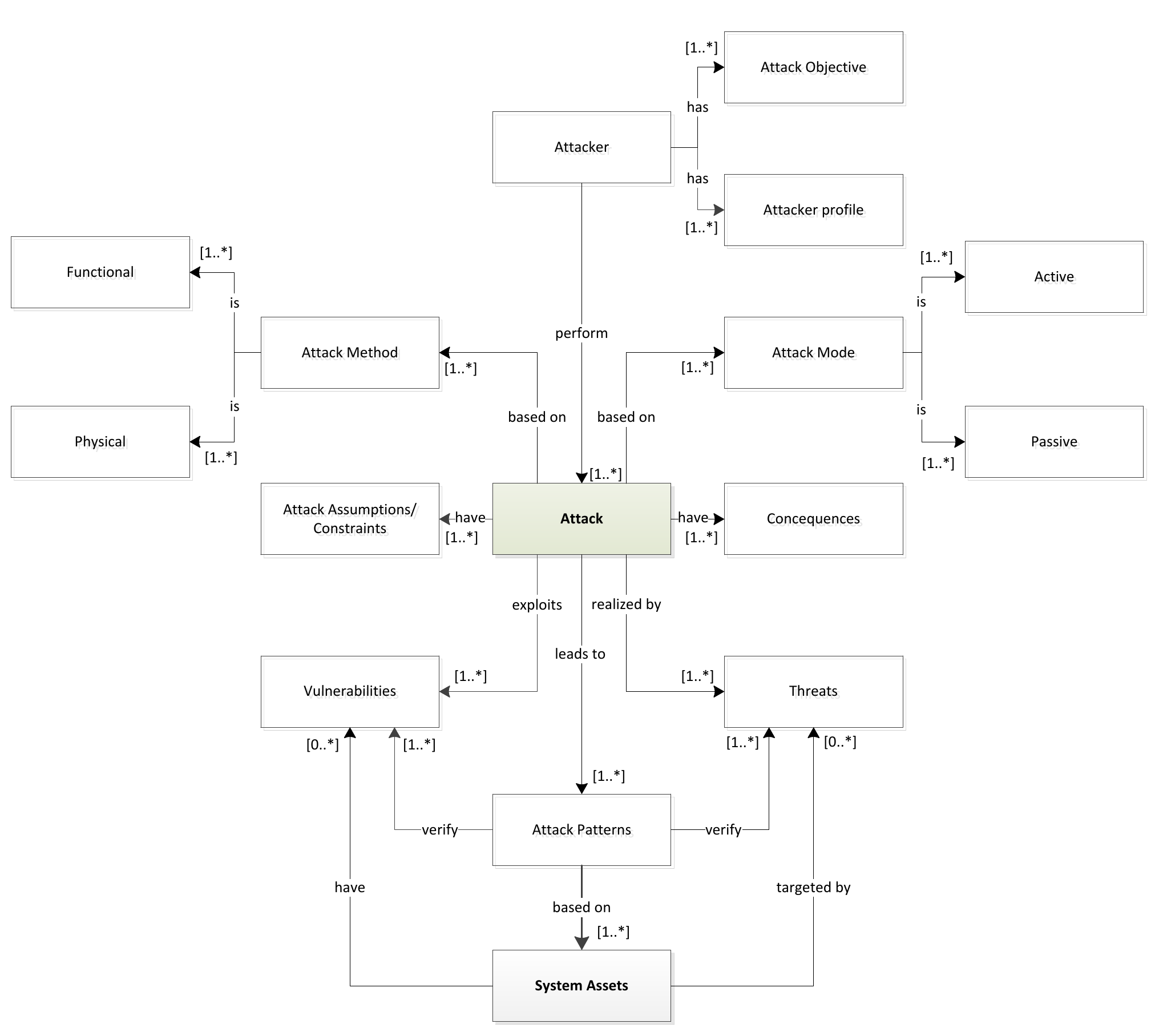}
        \caption{Security Attack Ontology}
    \label{fig:attack-tree-ontology}
\end{figure}

In this section, we define the security ontology -- Security Attack
Ontology -- for modeling different types of adversary's knowledge.
Security ontology constitutes a knowledge repository for capturing,
classifying, and sharing security related information. More
specifically, the definition of a security attack ontology aims at
building knowledge vocabulary for security attacks that could be
described including their type, mode, consequences, and such details
as described above. Figure \ref{fig:attack-tree-ontology} sums up our
analysis with respect to extracting different constructs and concepts
defined in well-known security standards (i.e., ISO/IEC 15408:2009,
ISO/IEC 18045, ISO/IEC 27000: 2012, ISO/IEC 17799:2005, NIST
SP-800:30, etc.) and security dictionaries (i.e., CVE, CAPEC, OWASP,
CLASP, etc.) in order to build the security attack ontology. This has
been modeled with the Ontology Web Language
\cite{DBLP:bibsonomy_dean03} using OWL classes. Our security ontology
use a flexible and easily extendable structure, which makes it
possible to seamlessly add new concepts.

\section{Attack Modeling}
\label{sec:attackmodeling}

The concept of security analysis is similar to the concept of
trade-off analysis in that there is also more than one way to attack
system assets, and an adversary may be trying them simultaneously or
just a subset of them. More precisely, an adversary can use distinct
attack paths or alternative approaches until reaching his attack
objectives. This is often illustrated through the attack trees, which
form a convenient way to systematically categorize the different.
Basically; attack trees (the term was introduced by Schneier
\cite{Schneier1999}) are multi-levelled diagrams consisting
of one-root, leaves, and children nodes. In addition, different node
values can be combined with AND, OR relationship to learn even more
about a system's security flaws and weaknesses. Specifically, the
purpose of an attack tree is to define and analyze possible attacks on
a system in a structured way. This structure is expressed in the node
hierarchy as well as in the form of logical operators (i.e.,
conjunctive (aggregation) or disjunctive (choice), etc.) for
expressing interrelationship between different attack tree nodes.
Thus, using both logical operators and node definition retains the
natural way security experts build the attack trees or fault trees
\cite{collaborativeattack,EvitaD2.3,Schneier1999,camtepemodelingand,Vignaattacklanguages}.
Actually, these two building blocks (nodes and logical operators) of
an attack tree can be modeled with the definition of constraint block
with a object functions and the part element of the parametric
diagram. Thus, at a conceptual level we can use parametric diagrams to
model attack trees. Let us present how we suggest representing attack
trees in SysML using the above-mentioned modeling constructs.

\subsection{Parametric Diagram} The parametric diagram is the second
new type of diagram introduced to describe constraints on system
properties to support engineering analysis. The parametric diagram is
a specialized variant of an internal block diagram that restricts
diagram elements to represent constraint blocks, their parameters and
the properties of block that they bind to. Parametric diagrams are
made up of one or more constraint blocks, zero or more part, and one
or more connectors \cite{SysML-Book}. The constraint block is used to
show which constraints are being used. The SysML specification
describes constraint blocks in terms of conditions that are
represented by mathematical equations. More precisely, the constraints
block contains an equation, expression or rule that relates together
the parameters given in the parameters block. The concepts behind
constraints can be extended to cover general rules that constrain
system properties and behavior such as authentication should be
performed BEFORE authorizing entity to access system resources, etc.
The use of a constraint block is called a constraint property and is
depicted on a parametric diagram. The interconnection between
constraint blocks and part or constraint blocks is shown on a
parametric diagram using zero or more binding connectors. Binding
connectors depict an equality relationship between the two connected
parameters or between a parameter and a value property. In the
parametric diagrams, a standard part element includes properties to
specify its unique identifier and text description.

\subsection{Attack Trees in Parametric Diagram}
 \label{subsub:attack-tree-modeling}
\begin{figure}[http] \centering \includegraphics[width =\columnwidth]{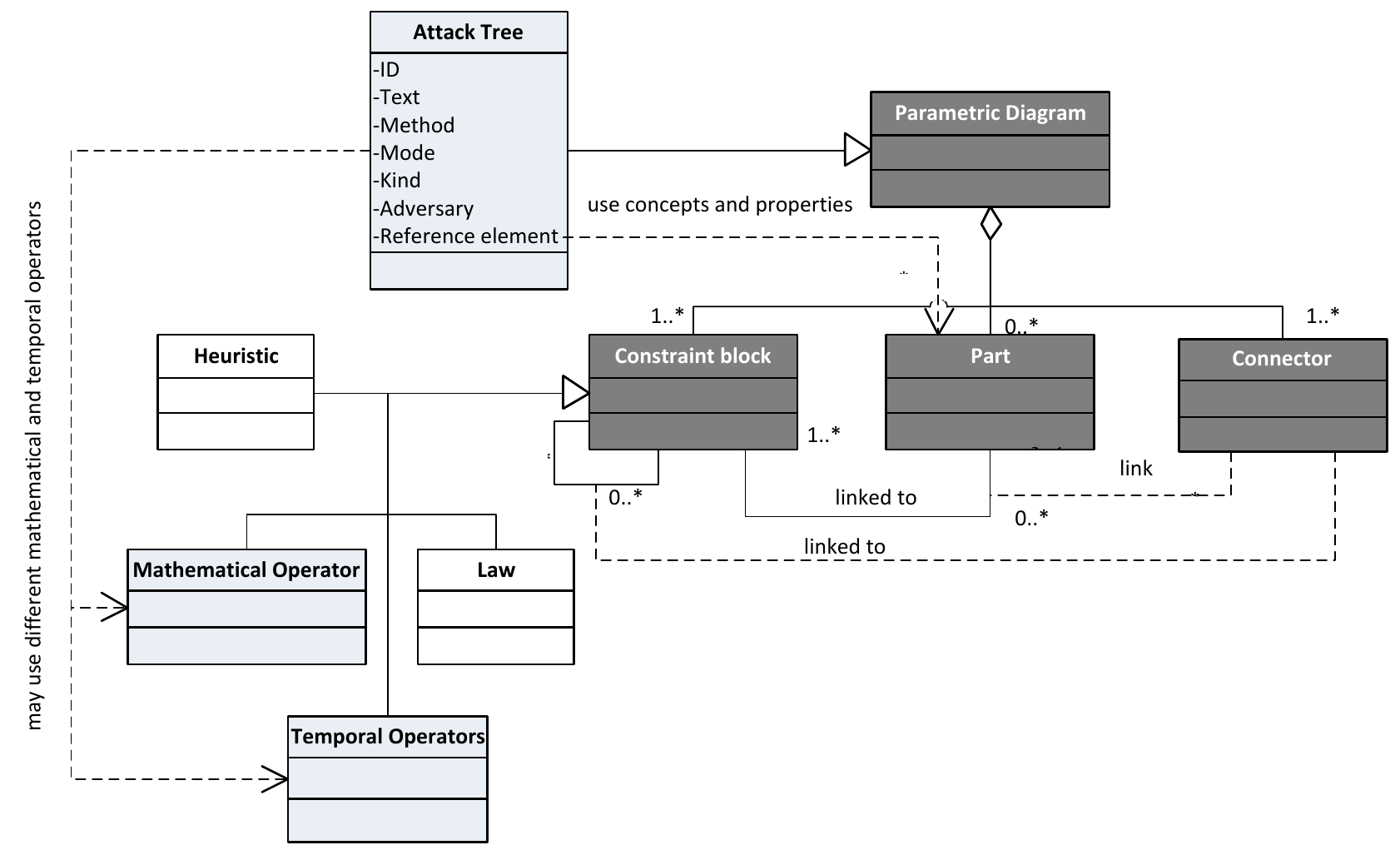}
		\caption{Metamodel for the SysML attack tree diagram}
	\label{fig:PD-metamodel}
\end{figure}

Let us first focus on the extension of the parametric metamodel (see Figure \ref{fig:PD-metamodel}) that is necessary for modeling attack trees.
Following the extension mechanism suggested in the SysML specification where the stereotype mechanism is defined to extend the existing  SysML classes, we create a new stereotype to represent security attacks: the  "attack tree" This is illustrated in Figure \ref{fig:PD-metamodel}.  As mentioned earlier, we mainly focus in this paper on expressing security knowledge to be shared and reused  throughout the system development process to design a secure system. In order to integrate attack related  knowledge, we extend the parametric diagram's "part" element with  ontological concepts and properties from the attack tree ontology, presented in Section \ref{subsec:so}.  We argue that such a representation is indispensable to precisely understand how attack trees can be manipulated during their construction and analysis. More details are given in Section  \ref{sec:sec4} about  the introduction of security reasoning into SysML models.
We use the  "constraint block"  element for the definition of set of constraints such as mathematical expressions (i.e., AND, OR, etc.) among the pieces of the security attack nodes. The objective of these operators is  to show the relationship among difference attack nodes.  More precisely,  we use OR operator to represent alternatives ways an adversary tries to achieve his attack objectives. For instance, an adversary has to perform either one of the attacks  "hijack  authenticated session" OR  "disconnect client"  to accomplish his attack goal.  AND relationship represent different steps toward achieving the same goal, for example,  by assuming an adversary can gain root access of vehicle Communication Unit (CU) if and only if  he can  tamper  the on-board communication unit.
In our attack tree modeling approach, rather than considering only these two types of logical operators,  we also consider  temporal operators (i.e., AFTER, BEFORE, SEQUENCE, etc.).  We in particular allow security experts to capture temporal dependencies between attack nodes and sequences in an attack. For instance, in order to install the bogus authority keys an adversary first have to switch an ECU into a re-programming mode. Furthermore, we can represent the ordering between attacks by using the SEQUENCE relationship. We use a "connector"  element to link  zero or more "part"  with constraint block.  The use of a "constraint block", "part", and "connector" element  for  building  attack trees is  shown in Figure \ref{fig:Attack-Tree-model}.
\subsection{Knowledge-Centric Attack Tree Modeling}
\label{subsec:kcat}

\begin{figure}[http] \centering \includegraphics[width
=\columnwidth, height = 2.25in]{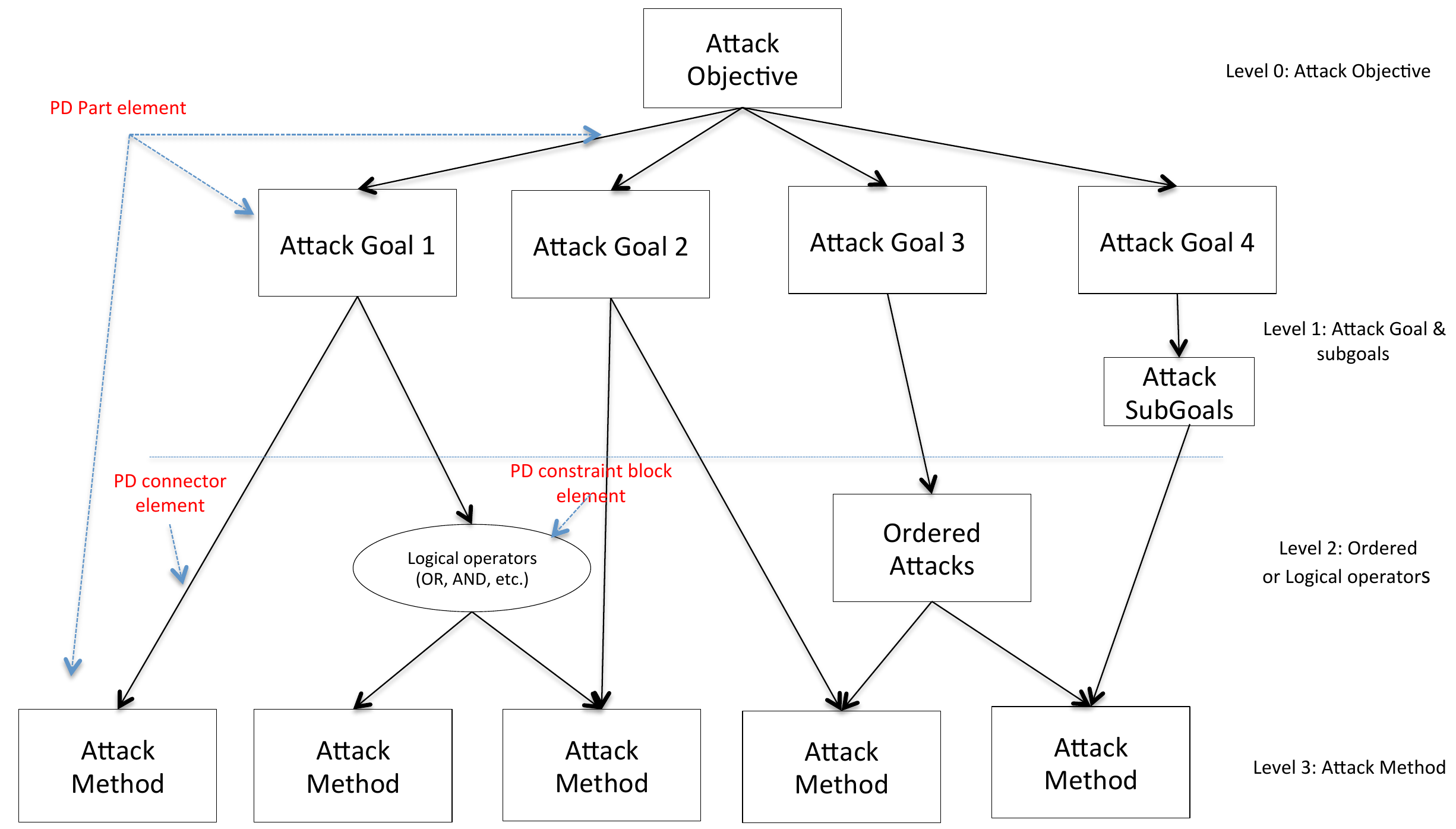}
		\caption{Generic knowledge-centric attack tree structure}
	\label{fig:Attack-Tree-model}
\end{figure}

 An overall procedure for attack tree modeling looks like this:
\begin{enumerate}
\item Build attack tree rooted (Level 0) on  an abstract
"attack objective". We use the "part" element to model each attack tree node.
\item  Its child nodes (Level 1) represent different "attack goals" that could
satisfy this attack objective.  Attack goals and attack objectives are linked via a binding "connector".
\item For each attack goal node:
\begin{itemize}
\item  Decompose into a number of "attack methods" (Level 3) that could be employed to
achieve the attack objective.
\item  Specify the logical relationships (Level 2) between different attack methods, if there are.  We use the "constraint block"  to specify these logical expressions. At this stage, we also consider intermediate steps that represent attack method at a certain level of abstraction.
\end{itemize}
\item The attack tree terminates when leaf conditions (basic operations are described that gives all details of the attack) are reached that meet the adversary's capabilities.
\end{enumerate}
The attack tree modeling approach that we advocate provides a bridge between the typical attack trees modeling approaches \cite{Schneier1999}, and the anti-goal models approaches \cite{VanLamsweerde2007}. More precisely, the first two steps of our attack tree modeling approach  are equivalent to the KAOS anti-goal model \cite{VanLamsweerde2007}, which provides the top down approach for modeling attacks.  The next two steps correspond to the standard attack tree modeling approach, where attacks are identified from bottom up perspective.  Figure \ref{fig:Attack-Tree-model} sums up these two approaches.

\subsection{Tooling}
The aforementioned approach has been applied with TTool \cite{TTool}. TTool is an open-source toolkit that supports several UML/SysML profiles, including TURTLE [3] , DIPLODOCUS [4], and AVATAR [27,18]. TTool includes diagramming facilities, formal and non-formal code generators, model simulators, model-checkers and analysis tools. DIPLODOCUS is focused on the design space exploration of complex embedded systems. TURTLE is now deprecated and replaced with AVATAR. the latter targets the design of embedded applications. The main strength of TTool is to hide knowledge of the underlying simulation or formal proofs techniques, thus offering a press-button approach to perform safety or security proofs.
We have used the SysML-Sec profile as a part of the AVATAR profile to build attack trees.

\section{Integrating Knowledge Bases and Reasoning into SysML}
\label{sec:sec4}
Syntactically, SysML and ontology languages (i.e., OWL, OIL, etc.) have a lot of similarities. While SysML makes use of a graphical formalism, it also aims at defining the semantics of a system with constructs like blocks, associations, part properties, and relationships between models and sets of model elements.
Ontology languages use classes, properties, relationships, and individuals as basic knowledge constructs.
For instance, OWL defines classes by appropriate and implicit logical constraints on properties of their subclasses and concepts.
The integration of both approaches enables engineers to add reasoning arguments to the explicit documentation of system models, and to define more precise relationships in the course of a typical model-based development process. We discuss in the following how ontologies and inferences on ontologies are used to enrich the SysML-Sec framework.
\subsection{Annotating SysML Diagrams with Ontological concepts}
\begin{figure}[http] \centering
   \includegraphics [width=\columnwidth] {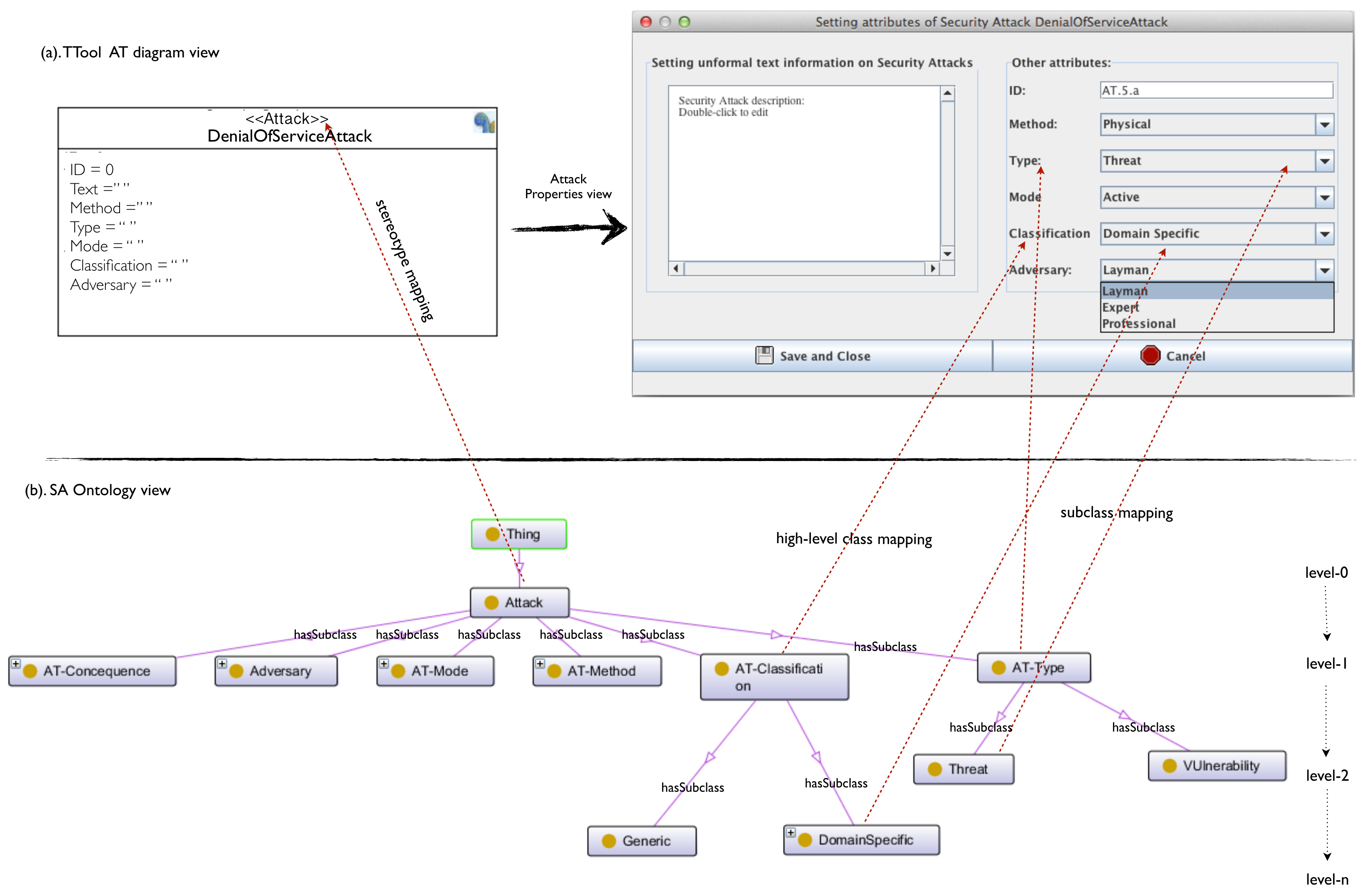}
        \caption{Mapping of the security attack ontology concepts into the SysML Attack Tree diagram}
    \label{fig:Ontology-Sysml-ttool}
\end{figure}

In this section, we focus on annotating security concepts and terms defined in the security attack ontology (cf. Section \ref{sec:MSP}) with attack tree diagrams. Figure \ref{fig:Ontology-Sysml}, presents an overview of an integration approach for embedding ontological concepts and terms into SysMLsec models. As previously stated in section \ref{sec:attackmodeling} we can add the ontological concepts and terms into SysML models by extending the SysML metamodel by including the user defined stereotypes or properties and tagged values. Let us first consider the Attack Tree diagram (presented in section \ref{subsec:kcat}), and map adversary  related ontological classes (cf. section \ref{subsec:so}) to the Attack Tree diagram. We build the integration approach based on three core ideas:
\begin{figure}[http] \centering
   \includegraphics [width=2.25in, height=2.85in] {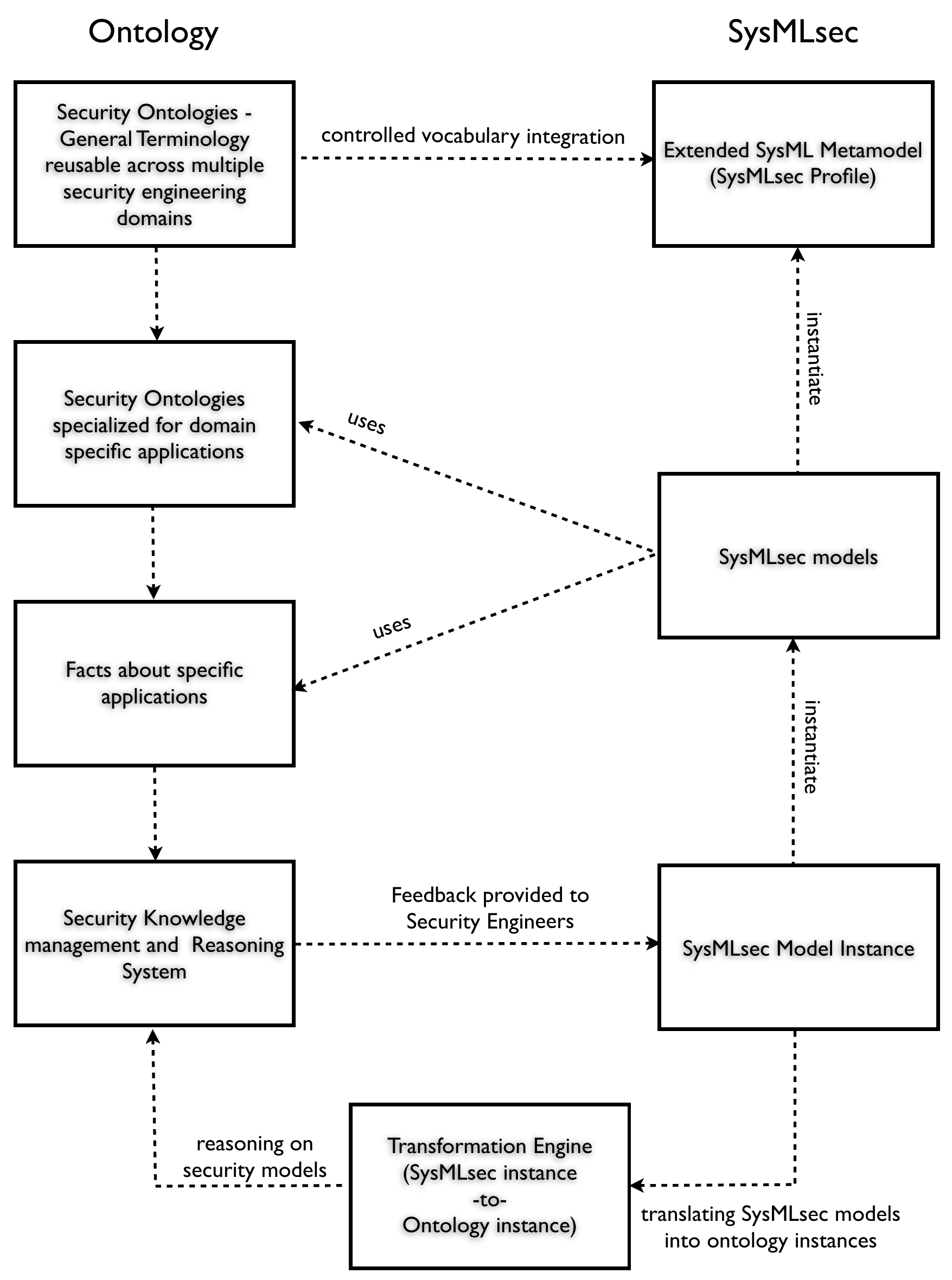}
        \caption{Integration of ontology reasoning on security with SysML}
    \label{fig:Ontology-Sysml}
\end{figure}
\begin{itemize}
\item We have defined the "Attack" stereotypes (see Figure \ref{fig:Ontology-Sysml-ttool}) to represent the security attack ontology in the SysML Attack Tree diagram.
\item We integrate high-level ontology classes (see Level-1 in Figure \ref{fig:Ontology-Sysml-ttool}.b) as a SysML Attack Tree diagram properties (i.e., type, kind, etc.) as shown in Figure \ref{fig:Ontology-Sysml-ttool}.a.
\item We use ontology subclasses (see Level-2/n Figure \ref{fig:Ontology-Sysml-ttool}.b), as tag values of the SysML attack tree diagram property element. This is illustrated in Figure \ref{fig:Ontology-Sysml-ttool}.a. These values constitute a controlled vocabulary. Thus, it provides a canonical set of mapping mechanism in order to deal with integration of ontological concepts into the SysML.
\end{itemize}

According to these rules, every Attack Tree diagram extended with an "Attack" stereotype is also associated with ontology concepts and terms as shown in Figure \ref{fig:Ontology-Sysml-ttool}.a. The diagram consists of two parts; standard SysML parametric properties (e.g., id, text) and  extended ontological properties (e.g., kind, type, classification, etc.). The discussion here will be limited to the extended attack's property constructions that can be directly translated to ontology classes. It can be seen that, for each high-level class of the security attack ontology (see Figure \ref{fig:Ontology-Sysml-ttool}.b.) we basically define a new property element in the attack diagram. This is illustrated in the Figure \ref{fig:Ontology-Sysml-ttool}.a. These properties are then populated with the subclasses and concepts defined in the security attack ontology as its tagged values. In particular, the properties and tagged values are specified in the same manner as the classes and subclasses concept described in the security attack ontology. For example the "Adversary" class takes Layman, Expert, and Professional as its tagged value. Accordingly, we map all other concepts and terms defined in the security attack ontology into the "part" element.

\subsection{Reasoning with SysMLsec Models}
\label{sec:example}

\begin{figure}[http] \centering
   \includegraphics [width=\columnwidth, height=2in] {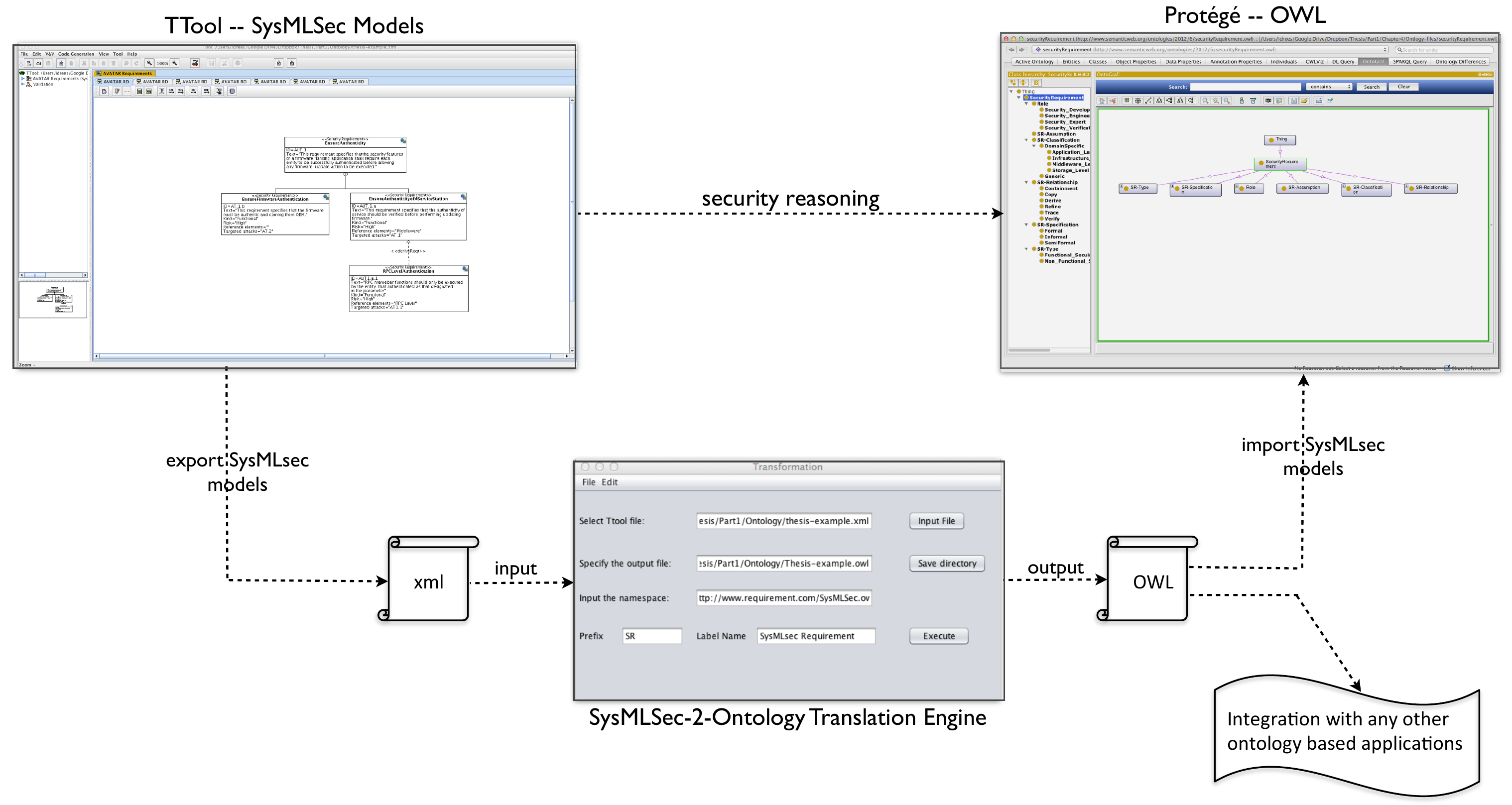}
        \caption{Integration of ontology reasoning on security with SysML}
    \label{fig:Ontology-Sysml-integration}
\end{figure}

In this section, we describe the extent to which we can use the capabilities of ontologies to reason about these different security concepts defined in Attack Tree diagram. In particular, our objective is to enable the security engineers to have access to various ontological concepts and terms, and to reason on these models. Although, with integration of ontology classes and subclasses into the SysML Attack Tree diagram, we already provided the partial reasoning capabilities to reason about different security concept within the SysML models. More precisely, when security engineer select a particular concept in the SysML diagram, for instance, Attack is a "Domain specific" threat, we annotate the structure of the sub-classification with the tagged values that belong to the domain specific class such as an application specific, middleware specific, etc, as shown in the property view of Figure \ref{fig:Ontology-Sysml-ttool}. In a similar way, for each ontology class we apply the same approach and limit the knowledge space for security engineers to specify only those concepts and terms that belong to the super class or the parent class. Thus, provide means to reason about security concepts within the SysML models, which brings additional power to the development of security models like consistency checking (i), concept satisfiability (ii), and concept classification. The shortcoming is that we cannot specify the reasoner calls in relation to one another or in relation with other security models such as security goals, attack, system architecture, etc., which is our core objective. In order to fulfil this design objective, we have implemented the "SysMLsec-to-Ontology" translation engine as shown in the Figure \ref{fig:Ontology-Sysml-integration}. The translation engine, we have implemented for mapping from SysMLsec models to the OWL description, contains a set of rules that match security constructs and transform them into equivalent instance of ontology.

The primary purpose of this translation engine is to make the security engineers able to reason about their security models using well known and  efficient reasoning engines such as SPARQL \cite{prudhommeaux2006-1}, OWL-QL \cite{Fikes03owl-ql}, RACER \cite{Haarslev03racer:an}, SQWRL \cite{conf/semweb/OConnorD08}, etc. Engineers can directly make use of reasoning capabilities of these engines within the context of current engineering practice and tools without building and using some separate ontological models. In particular, our objective is to give the security engineering  a more precise way to employ reasoning in the course of a typical model-based development process.  On the other hand, since  we obtain an OWL described document, we can integrate our SysMLsec models with any other  ontology based applications such as integrating the security requirement knowledge with security resource annotating approaches like \cite{DBLP:conf/otm/KimLK05}, or providing input to the ontology based risk analysis approaches \cite{4076691} in order to compute risk metrics.

In this paper, we refer to SQWRL (Semantic Query Web Language) \cite{conf/semweb/OConnorD08} as a query language because of its  concise, readable, and semantically
robust semantic. SQWRL is a SWRL-based \cite{Horrocks04} query language that can be used to query OWL ontologies and provided in
Prot\'{e}g\'{e} 4.2 beta version \cite{protege}. To retrieve knowledge from OWL ontology, SQWRL provide SQL-like operation. The form of rule is
  \begin{eqnarray}
 antecedent \rightarrow consequent\nonumber%\label{eq:sample-eq}
\end{eqnarray}
In this rule an antecedent part is referred to as the body, and a consequent part is referred to as the head. Both the body and head consist of positive conjunctions of atoms: \begin{eqnarray}
  Atom\; \wedge \;Atom\;  \rightarrow Atom \wedge \; Atom\nonumber
\end{eqnarray}
This rule can be read as if all the atoms in the antecedent are true then the consequent must also be true. Here, an atom is an expression of the form P(arg$_{1}$, arg$_{2}$, ...arg$_{n}$), where p is a predicate symbol and arg$_{1}$, arg$_{2}$, ..., arg$_{n}$ are the terms or arguments of the expression. In our approach, the predicate symbols can include security ontology classes (i.e., asset,  goal, attack,  security requirement, etc.), properties (i.e., hasFunction, hasSequence, hasAvoidGoal, etc.) or data types. Arguments can be class individuals (i.e., type, classification, adversary, etc.) or data values, or variables referring to them. In the further course of this thesis, we will use the above-mentioned  SQWRL query expression to retrieve, manipulate, and reason about different security-related information.

\section{Related Work}
\label{sec:SOA}
\input{SOA}

\section{Conclusion}
\input{Conclusion}

\bibliographystyle{eptcs}
\bibliography{bibtex-dbdec}

\end{document}

%% file: Introduction.tex
 Security attack, whatever objective it has, is deontology that derives the
wrongness of one's conduct which compromise the security objectives. Frequent reports about security vulnerabilities show that still many deficits
 exist in the development of secure software systems. The problem
 is even more pressing as the adversary activity and the destructiveness of
 attacks  have increased over the last years. It is generally agreed that a  central problem in the design of secure systems and the security analysis of existing systems is the danger of overlooking the system from particular standpoints  \cite{collaborativeattack}.  This corresponds  to situations in which security of the system  is analyzed and described in terms of making the system  secure by preventing weak links.  
 In such a context, it is not sufficient to discover security attacks only at
overlooked  weak point of the system; there is also a need to analyze the
information flow control issues,  especially when the underlying platforms and
infrastructures are also made of services themselves. 
Security analysts also need to consider threats to these
underlying infrastructure and middleware for a particular security realization,
as the assets to be protected originate both from the horizontal (i.e., between different entities and components) and vertical (i.e., multiple layers)
compositions.

% That is, the coordination of multiple architectural layersover  which the application/services is deployed, including operating systems, application servers, middleware services, hardware, etc.\\
% The security that may mitigate potential threats to these assets have to be
% deployed at different parts of that stack, and in a coordinated manner.
A related problem is that it is easier to analyze the protection level at each
separate layer in the system architecture stack,  but become vulnerable to various
security exploits and flaws in a coordinated manner \cite{EvitaD4.4.2,EvitaD2.3,collaborativeattack}.
Because of their complexity and of the varying degrees in which system assets are
deployed and executed, it is often the case that a system is compromised through
a path its developers never have thought of. 
What is worse, a local security
attack and vulnerability or a mismatch between the security  mechanisms adopted
at different locations can have dire consequences, potentially putting  the security of large
system at stake. Most of such security attacks stem from the limited knowledge shared
between various security-engineering activities
 that collaborate with each other  and the expression
of  their interdependencies. \\
One thing is that it is not easy to discover all
parts of a system that are relevant for its security. In mainstream practice,
this knowledge is often spread across different architecture layers, and correspond to various system
development activities such as  system architecture design, goal specification,
%security requirements, testing, etc. 
%
%Meanwhile,  it is unlikely that the same
%set of domain knowledge, assumptions,  or sharing of a common understanding of information among people (i.e., security experts, developers, etc.), which typically do not belong to the
% same group,  will fit altogether in a systematic and intuitive way.  This is already evident in the form of having several
% security standards (i.e., ISO/IEC 15408:2009, Six Sigma, Octave, etc.) and
% security dictionaries (i.e., CVE, CAPEC,  OWASP, CLASP, etc.) in the field of
% security threats and vulnerability analysis, where similar concepts are specified
% and treated distinctively. 
In general, for a thorough security evaluation, one needs to take into account 
these different  knowledge perspectives. In this context, in this paper, we aim at proposing a  security analysis model derived from the conceptual constructs of security ontologies that will serve as the common knowledge repository for discovering, analyzing, and sharing attack knowledge with other system development activities. Thus, it will offer means to analyze the security of the system in such a way that it is possible to discover simple and complex security attacks and vulnerabilities at different levels of system abstraction. 
Furthermore, the concept of attack tree, modeled in SysML Attack Tree Diagrams, is brought in as the foundational graphical representation for modeling and embedding the collected security attacks knowledge into the security attack ontology. In this manner, the attack trees are completely parameterized by the ontological concepts so that it is possible to handle simultaneously several knowledge bases associated with security attacks and vulnerabilities. In particular, the knowledge based attack trees ease the process of keeping security attack specifications clear and understandable, minimizing the inconsistencies and helping to achieve maintainability -- even when security attacks are drafted cooperatively by several entities as well as at different system development stages. 

%% file: MSP.tex
In IT security engineering, to be on the safe side, we must assume that each attack scenario that is possible and promises whatsoever small benefit will definitely be carried out by someone. In this regard we first need to make a clear distinguish about: what does an adversary look like in distributed systems when different entities are involved such as when a client is a user, an owner, and a service provider and when some or all of the entities in the system can become adversaries?, Is there a hierarchy of adversaries when attacking such heterogeneous systems?, and so on. More specifically, security attacks are hard to understand, often expressed with unfriendly and limited details, making it difficult for security experts and for security analysts to create intelligible security specifications. For instance, to explain "Why" (attack objective), "What" (i.e., system assets, goals, etc.), and "How" (attack method), adversary achieved his attack goals. We introduced security attack ontology  by taking into account security standards and security dictionaries and deriving the features, in terms of classes and sub-classes that were needed in such situations. Security attack ontology has been designed to enable the specification of security attacks in a concise, readable, and extensible way. Following, we detail different types of knowledge that an adversary require or use to perform an attack.

\subsection{Adversary Profile}

The adversary profile depicts the attack potential that is a measure of the minimum effort to be expended in an attack to be successful. In ISO/IEC 15408:2009 the attack potential is defined as a "measure of the effort to be expended in attacking a TOE, expressed in terms of an adversary's expertise, resources and motivation". Essentially, the attack potential for an attack corresponds to the effort required creating and carrying out the attack. The higher the adversary's motivation is the higher efforts they may be willing to exert. After having performed a comparative analysis of several security specifications and standards, we suggest the following abstract level taxonomy (see Figure \ref{fig:adversaryProfile}) to be considered during an analysis of the attack potential:
\begin{figure}[http] \centering
   \includegraphics [width=\columnwidth] {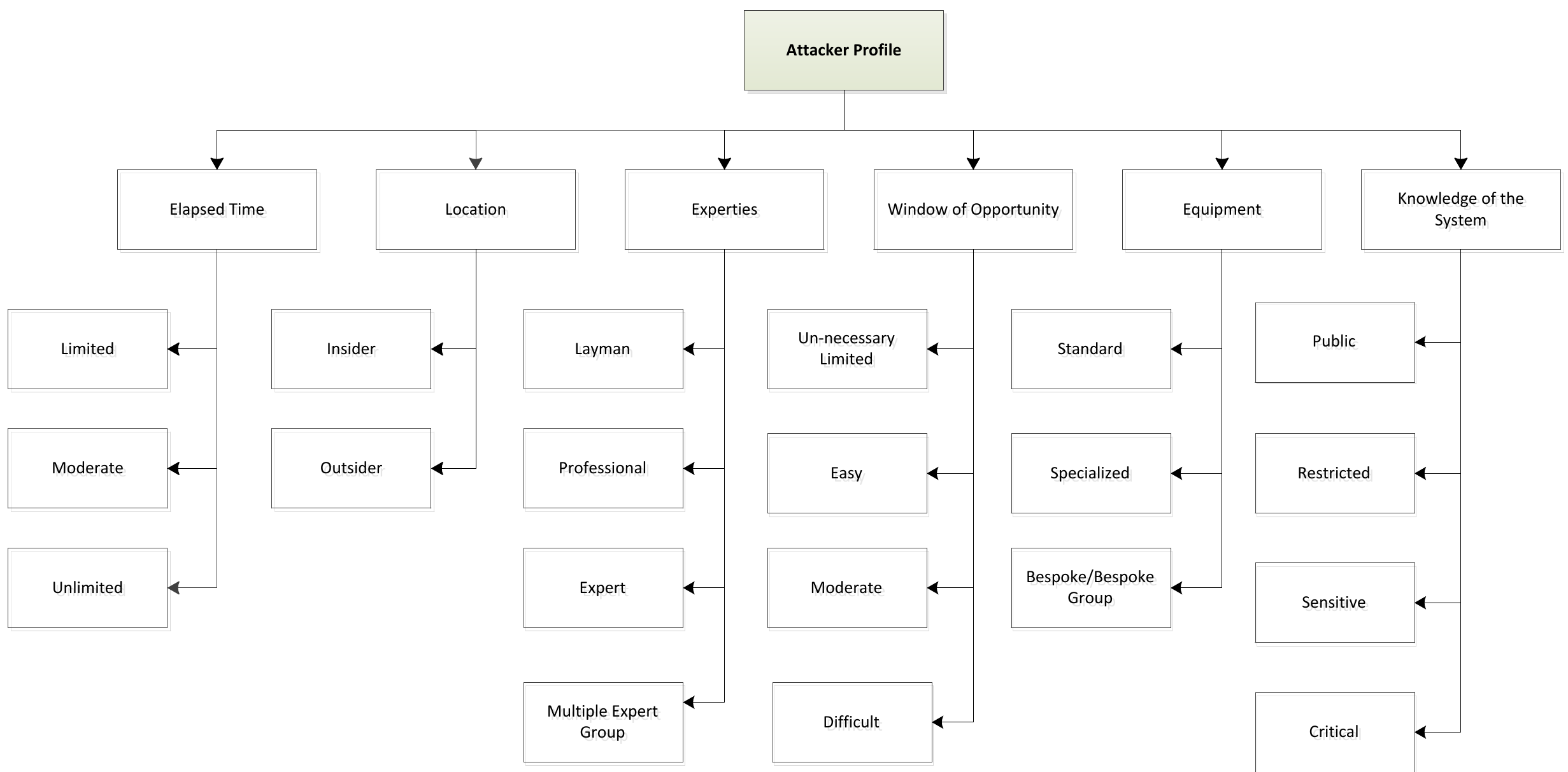}
        \caption{Adversary Profile Taxonomy}
    \label{fig:adversaryProfile}
\end{figure}

\begin{itemize}
\item \textbf{Elasped Time:} This is the total amount of time taken by an adver- sary to identify that a particular potential vulnerability may exist, to develop an attack method and to sustain the effort required for mounting the attack.
\item \textbf{Expertise:} This refers to the required level of general knowledge of the underlying principles for mounting an attack (i.e., system archi- tecture, security components, etc.), product types or attack methods.
\item \textbf{Location:} This refers to the knowledge and the capabilities, which an attacker may have, depending of his/her location; this is typically reflected by the terminology for an Insider or Outsider attacker. For instance, insider attack agents are likely to have specific attack objectives, potential, and have legitimate knowledge and access to the system.
\item \textbf{Window of opportunity:} This concept has a relationship with the elapsed time factor. Identification and exploitation of vulnerability may require considerable amounts of accesses to a system that may increase the likelihood of detection of the attack. In contract, some attack methods may require considerable effort off-line, and only brief access to the target to exploit.
\item \textbf{IT hardware/software or other equipment:} This refers to the equip- ment required to identify and exploit vulnerability.
\item \textbf{Knowledge of the system under investigation:} This refers to the spe- cific expertise required in relation to the system under investigation. Though it is related to general expertise, it is distinct from it.
\end{itemize}

\subsection{Adversary Objectives}
Attack objective suggest particular types of adversary and his capabilities, as well as associated attack motivation. At the abstract level of specification attack motivations can be broadly categorized as:
\begin{itemize}
\item \textbf{Individual Benefits:} Personal advantages can be gained in different ways and for different purposes. For instance, gain reputation as hacker, financial gain fraudulent commercial transactions, etc.
\item \textbf{Economical Benefits:} These motivations and underlying objectives should be envisaged at an organizational scale.
\item \textbf{Political Benefits:} The main goal of the attacker is to destroy the reputation of an organization or an individual system asset. For example, acquiring system design information or for the purposes of fraud, industrial/state espionage or sabotage.
\item \textbf{Criminal Benefits:} An augmentation of the attack motivation to harm an individual for the purposes of criminal or terrorist activity, destroy or financial harm, destructive attacks or intellectual property attacks, etc.
\end{itemize}
\subsection{Attack Mode used by an Adversary}
The attack mode refers to the actions that an adversary takes during the execution of an attack and that can be labeled as active or passive attacks:
\begin{itemize}
\item \textbf{Active Attacks:} modifying the behavior of the system.
\item \textbf{Passive Attacks:} aiming at information retrieval without modifying the behavior.
\end{itemize}

\subsection{Attack Method used by an Adversary}
The attack methods are related to the attack mode class. The attack method can be classified into either functional (logical) attacks or physical attack methods:
\begin{itemize}
\item \textbf{Physical Attacks:} Attacks physically modifying the behavior of the system. 
\item \textbf{Functional Attacks:} From the functional point of view, attacks aiming at logical manipulation of information
without physically modifying the system behavior.
\end{itemize}
\subsection{Attack Consequence}
Attack Consequences refers to an impact of security breach or outcomes that are not the ones intended by a purposeful system action. The attack consequences can be classified as:

\begin{itemize}
\item \textbf{Usurpation:}  is a derogatory term used to describe either a misappropriation or misuse of the system functionalities.
\item  \textbf{ Disruption: } is an event that causes an incapacitation, corruption, obstruction, and unplanned deviation from the expected system behavior, according to the functional and non-functional objectives.
\item  \textbf{Deception:} is defined as masquerade, falsification, and repudiation actions taken by an adversary, to thereby causes a system to accept as true a specific incorrect version of reality.
\item \textbf{Disclosure:} enables an adversary to gain valuable information about a system and its functionalities either by exposure, interception, inference, intrusion, etc. that tries to uncover the details of a system.
\end{itemize}

%% file: SOA.tex
In the following we will introduce different threat modelling profiles. These modelling profiles capture a certain types of information and results in different types of threats and vulnerability models and security design solutions. A number of extensions of UML (i.e., UMLsec \cite{Juerjens:2003:SSD:996198}, Anti-Goal  \cite{VanLamsweerde2007}, Misuse cases \cite{891363},  Abuse cases \cite{Lin:2003:IAF:942807.943895}, etc.),  allow to express security relevant
information within the diagrams in a system specification have been proposed. For instance, Abuse Frames are based on the Jackson's problem frames approach \cite{Jackson:2000:PFA:513720} and is intended to analyze security problems in order to determine security vulnerabilities and to derive security requirements. This approach introduces the notion of anti-requirement (similar to the concept of an anti-goal \cite{VanLamsweerde2007}) to describe the behavior of a malicious user that can subvert an existing requirement. The basic idea behind the definition of abuse frames is to bind the scope of a security problem with anti-requirements in order to derive security requirements. Such explicit and precise descriptions facilitate the identification and analysis of threats, which in turn drive the elicitation and elaboration of security requirements. Possible ways of misusing system functionality can be specified by an extension of UML. Misuse case diagrams not only shows regular actor/use case relations but also can model threats that threaten use cases, and countermeasures that mitigate these threats. Misuse cases extend the traditional use case approach to also consider misuse cases, which represent behavior not wanted in the system to be developed. A misuse case diagram contains both, use cases and actors, as well as misuse cases and misusers. Development of misuse cases allows the identification of security attacks and associate security requirements during application development. In \cite{Whittle:2008:EMC:1368088.1368106}, misuse cases are further analyzed and author's presents a formal representation of misuse cases and provide an intuitive way to executable misuse case model. Although misuses cases are not entirely problem-oriented as they represent aspect of both problems and solutions, they have become popular as a means of representing security concerns in the early stages of software development. Yet, to best of our knowledge, none of them provides the expressivity required to deal effectively with system-wide security attacks. Another major group of contributions to the conceptual modeling of security attacks like KAOS \cite{Lamsweerde03fromobject} and Secure Tropos \cite{Mouratidis02anatural}, etc., have defined their own graphical formalism each of which allows to express security relevant information (i.e., goal, anti-goals, requirements, obstacles, etc.). However, security issues involve special concerns that these traditional software engineering languages do not consider. Consider, for example, a general behavior modeling notation that expresses interactions of entities in the system without considering the harmful behavior of an adversary. Thus, the models do not convey the impacts of the malicious behavior of the adversary on requirements, design, and architecture to the next phases of system development lifecycle.

%% file: Conclusion.tex
Given an input for our knowledge centric design methodology, the security analysis process helps to both classify identified attacks, but also to think about new ones, given a category. Knowledge centric attack tree is a combination of both top-down and bottom-up approach to provide a support tool to security analysts. The purpose of developing the ontology driven attack trees is to identify possible security threats and to allow aspects such as the desirability (to the adversary), opportunity, probability and severity of attacks to be assessed in order to share knowledge among various system development activities (i.e., security requirements engineering, protocol design, testing, etc.). We believe that, on the one hand, ontology based security analysis is expressive enough to describe several real-world security attacks with a multi faceted approach; at the same time, it provides constructs to map and relate security attacks with other system development activities.